\documentclass[aps,preprint,
superscriptaddress
]{revtex4-1}

\usepackage{graphicx,subfigure}
\usepackage{dcolumn}
\usepackage{epsfig}
\usepackage{latexsym}
\usepackage{amsfonts}
\usepackage{amssymb}

\begin{document}

\title{Directing liquid crystalline self-organization of rod-like particles through tunable attractive single tips}%

\author{Andrii Repula,$^1$ Mariana Oshima Menegon,$^2$ Cheng Wu,$^1$ Paul van der Schoot,$^{2,3}$ and Eric Grelet}%
\email[]{eric.grelet@crpp.cnrs.fr}
\affiliation{
Centre de Recherche Paul-Pascal, CNRS \& Universit\'e de Bordeaux, 115 Avenue Schweitzer, F-33600 Pessac, France\\
$^2$ Department of Applied Physics, Eindhoven University of Technology, PO Box 513, 5600\,MB Eindhoven, The Netherlands\\
$^3$ Institute for Theoretical Physics, Utrecht University, Princetonplein 5, 3584\,CC Utrecht, The Netherlands\\}

\date{\today}

\begin{abstract}
	
\textbf{Compelling justification:} Hard-core repulsion is the simplest interaction in Nature 
yet it 
drives the self-organization of many complex fluids. To investigate how enthalpy impacts upon entropy-dominated liquid crystalline states, we introduce a highly localized and tunable directional attractive interaction (or ``patch") on one of the tips of rod-shaped colloids. Our experiments and computer simulations show that increasing the patch attraction dramatically stabilizes the lamellar phase, a structure desired in materials science due to its outstanding mechanical and optical properties. Our work demonstrates that introducing patches in anisotropic nanoparticles adds to the control of their self-assembly.
	
\textbf{Abstract:} 
Dispersions of rod-like colloidal particles exhibit 
a plethora of liquid crystalline states, including nematic, smectic A, smectic B, and columnar phases. This
phase behavior can be explained by presuming 
the predominance of hard-core volume exclusion between the particles. We show here how the self-organization of rod-like colloids can be controlled by introducing a weak and highly localized directional attractive interaction between one of the ends of the particles. This has been performed by functionalizing the tips of filamentous viruses by means of
regioselectively grafting fluorescent dyes onto them, resulting in a hydrophobic patch whose attraction can be tuned by varying the number of bound dye molecules. We show, in agreement with our computer simulations, that increasing the single tip attraction stabilizes the smectic phase at the expense of the nematic phase, leaving all other liquid crystalline phases invariant. For sufficiently strong tip attraction the nematic state may be suppressed completely to get a direct isotropic liquid-to-smectic phase transition. 
Our findings provide insights into the rational design of building blocks for functional structures formed at low densities.

\end{abstract}

\maketitle

There is a considerable interest in the self-organization of fluid dispersions of nanoparticles into hierarchical structures and morphologies. On the one hand, there is a fundamental interest in elucidating the physical principles that govern the self-assembly of colloidal particles \cite{Glotzer2007}. On the other hand, there is also a technological interest in the context of fabricating novel functional materials bottom-up, that is, via self-assembly \cite{Grzelczak2008,Mason2018}.
For both reasons, anisometric building blocks are seen as highly promising systems, because of their versatility in surface functionalization and their ability to form complex architectures, as the liquid crystalline phases \cite{HZhang2009,Chaudhary2014,Gao2015,Gao2018,Steinmetz,Lee2009,Gibaud2012,Nakata2007,Manolis2016,NYSTROM2018}. 

Among the desired organizations relevant in the context of materials science and nanotechnology, layered structures stand out for their outstanding optical and mechanical properties \cite{Gabriel2001,Butler2013,Kahl2017,Lee2017}. Such lamellar or smectic phases usually appear at relatively high packing fractions, which tend to render them difficult to handle experimentally. 
\cite{Dogic2001,Kuijk2012,Lee2017}. It would therefore be appealing to develop methods and approaches to obtain the smectic ordering at lower particle loadings.
We have recently shown that the self-organization and phase stability of highly ordered liquid crystalline states of filamentous viruses, including the smectic phases, is dominated by volume exclusion and hence by entropy \cite{Grelet2014}, confirming the role of model colloidal system of these biological rods. 

Here, we go beyond relying on a purely hard-core interaction and homogeneous surface functionalization \cite{Zhang&Grelet2013,Zhenkun2018}, and introduce a tunable localized directional attraction between the tips of the virus particles by specifically grafting hydrophobic fluorescent dyes to one of the two ends of our virus-based colloidal rods. We investigate experimentally the impact of this ``enthalpic'' patch on the self-assembly behavior of the particles and compare our findings with computer simulations.
The regioselective functionalization of the tips of the rods into hydrophobic patches gives rise to highly localized attractive interactions, which strongly influence the relative stability and structure of the various liquid crystalline phases. In particular, we show how an increasing tip attraction stabilizes the smectic A phase at the expense of the nematic and eventually also the isotropic phase, extending the stability of the smectic A phase to relatively low concentrations. We demonstrate in this Letter the efficiency of introducing a single attractive patch in the design of anisotropic building blocks to sensitively control the balance between entropy and enthalpy, and thus to control the self-organization of these particles into the desired architecture.

\begin{figure}
	\includegraphics[scale=.12]{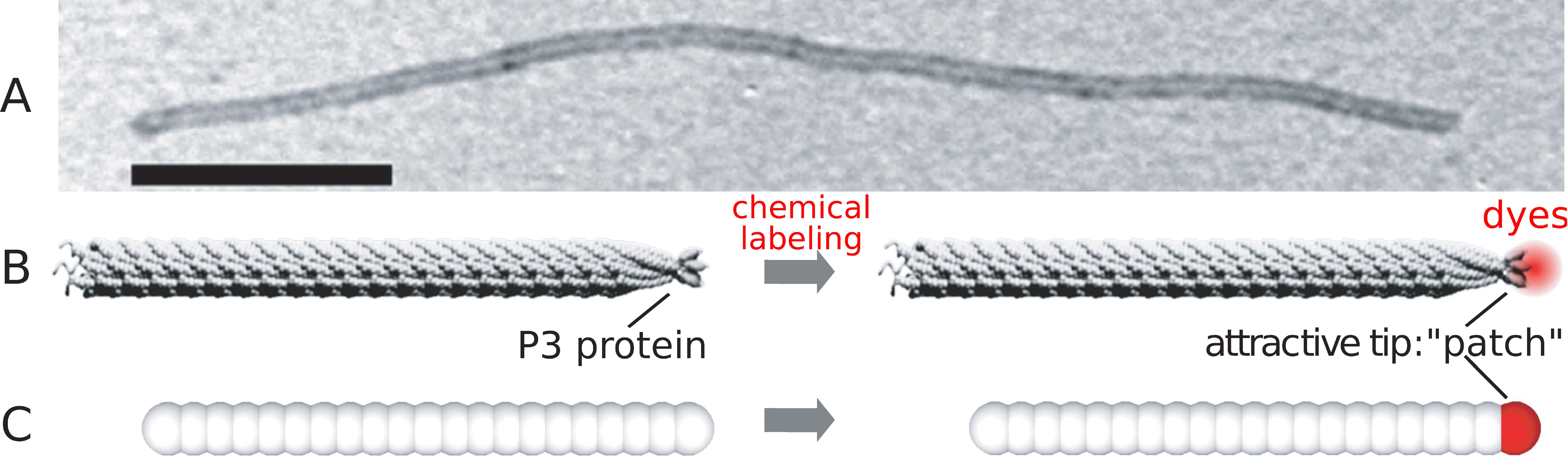}
	\caption{\label{Virus}\textbf{(A)} Transmission electron microscopy of the filamentous virus used in this work (Scale bar: 200~nm), and \textbf{(B)} schematic representation of pristine (left) and single-tip functionalized viral rod with red dyes (right), resulting in a localized directional attractive interaction. \textbf{(C)} Schematic of the semi-flexible rod-like particles, modeled as a bead-spring chain, used in computer simulations. The white beads from different particles interact via a repulsive soft-core potential, while the red ones located at one end of the rods are attractive.
	}
\end{figure}

In our experiments, we made use of  mutants of the filamentous bacteriophages 
M13KE and M13C7C, which only differ by the number of cysteine groups available at their proximal end on the P3 proteins (Fig. \ref{Virus}A-B). 
Both viruses are rod-shaped with a contour length of $L=1\,\mu{m}$, and a diameter of $7\,nm$. 
The particles are semi-flexible with a persistence length of $L_p\simeq 3L$ \cite{Barry2010}. 
The presence of cysteine residues only at one of the ends allows, after chemical reduction, for their selective bioconjugation with maleimide activated fluorescent compounds (Dylight 550 and 594 Maleimide, ThermoFisher), as described elsewhere \cite{delaCotte2017,Repula2018}. This results in single-tip labeled viruses, whose degree of functionalization, i.e., the average number of fluorescent dyes per virus can be controlled in our experiments from $n_\mathrm{dyes}=1$, 3 to 10 by varying the molar excess during the labeling reaction (See the Supplemental Material \cite{SM}.)  

The dye molecules are partially hydrophobic due to the presence of aromatic rings \cite{Zhang&Grelet2013}, implying that the number of grafted dye molecules dictates the size of the hydrophobic patch on the otherwise hydrophilic surface of the virus. It is reasonable to presume that the strength of the attraction between the virus tips increases with the patch area.
Whether there is a linear relationship between the number of dyes and the strength of the attraction is contentious, as the cysteine reduction and dye labeling leads to partial unfolding of the P3 tip proteins. This causes hydrophobic moieties of buried amino acids to become exposed to the aqueous solution. Still, it seems reasonable to assume that the number and size of these exposed hydrophobic groups increases with the degree of labeling, as confirmed by our experiments (See the discussion below).

Samples of single-tip functionalized virus suspensions have been prepared by dilution with BisTris-HCl-NaCl buffer, setting the pH at 7 and the ionic strength at 20~mM.   
These are then studied by optical microscopy \cite{Repula2018} and small angle X-ray scattering (SAXS, \cite{Grelet2014}) (see the Supplemental Material \cite{SM}).
In our molecular dynamics simulations, we model the chiral filamentous virus particles as achiral overlapping bead-spring chains, where 21 beads are connected via springs of rest length measuring half bead diameter and very large spring constant (Fig. \ref{Virus}C). Therefore, the aspect ratio of the simulated particles is 11, which is smaller by about one order of magnitude than the effective (i.e. accounting for the electrostatic repulsion between the charged viruses \cite{Grelet2014}) aspect ratio of the experimental particles. 
The consequences for the comparison between results from experiments and simulations are discussed below.
The beads interact via a steeply repulsive potential. A bending potential has been introduced to mimic the flexibility of the virus particles in order to reproduce the ratio between the persistence and contour lengths of the virus, $L_p/L\sim3$.
One of the end beads (displayed in red in Fig. \ref{Virus}C) representing the labeled virus patch interacts attractively through a Lennard-Jones potential with the other tip beads, and with a purely repulsive interaction with the other beads forming the rod particles. The strength of the tip attraction $u$ is the depth of the Lennard-Jones potential. 
Approximately 4600 chains are placed in size-adjustable simulation box initially organized in 8 AAA-stacked bilayers. We performed NPT simulations at various pressures,
using the simulation package LAMMPS according to a method described in \cite{Braaf2017}.

\begin{figure*}
 \includegraphics[scale=.3]{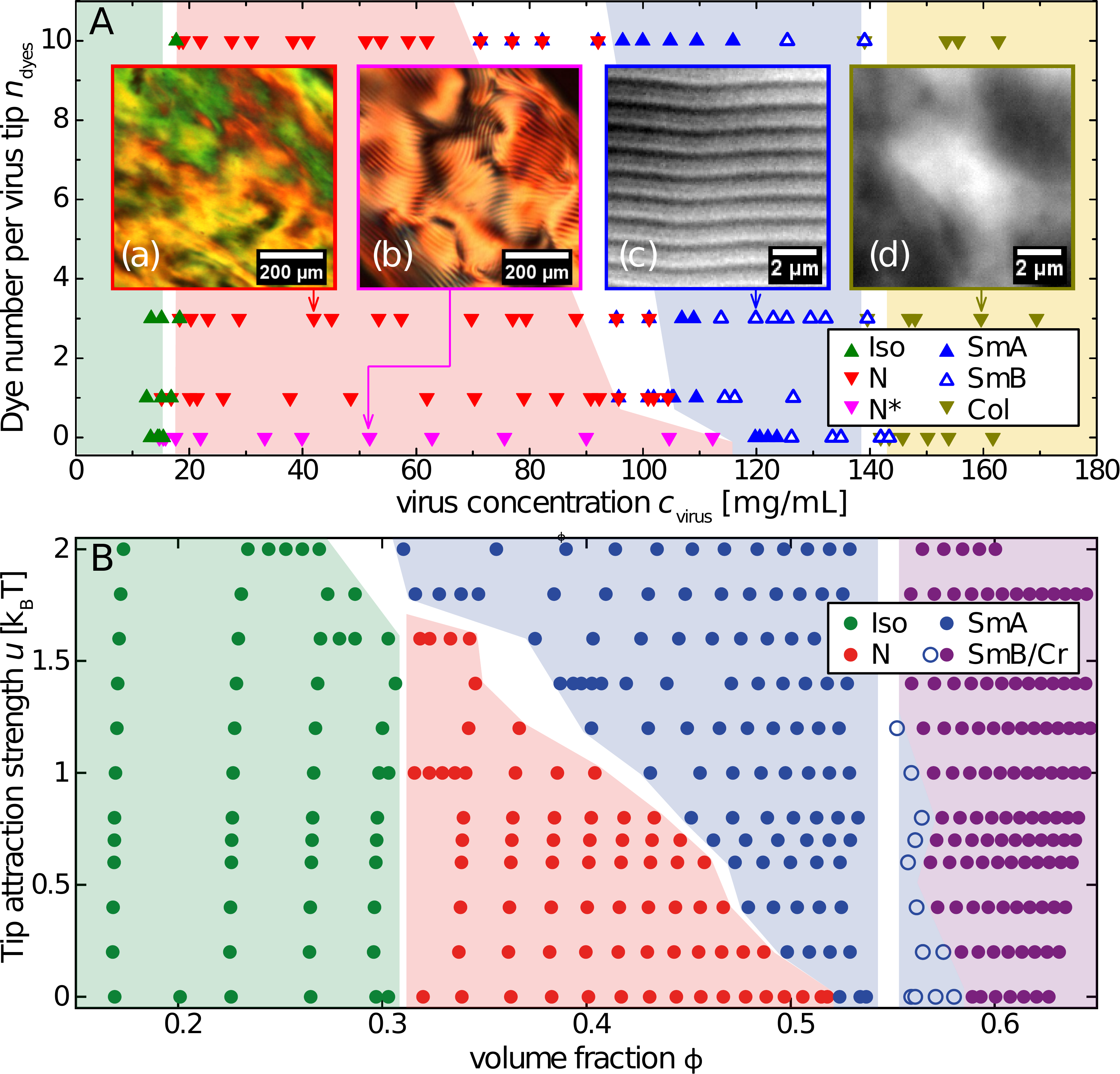}%
 \caption{\label{phaseDiagram}\textbf{(A)} Phase diagram of tip-functionalized viruses as a function of the mean number of grafted dyes per rod $n_\mathrm{dyes}$ and the concentration $c_\mathrm{virus}$.
 The system with 0 dyes corresponds to pristine (``raw'') viruses and the systems with $n_\mathrm{dyes}=1$, 3, and 10 dyes represent the tip-functionalized, patchy particles. The isotropic liquid (Iso) and (chiral) nematic (N$^{(*)}$) phases (Images (a) and (b)) have been identified by polarizing microscopy. The layered texture observed by differential interference contrast (DIC) microscopy in the image (c) is characteristic of the smectic A (SmA) and B (SmB) phases, and it lacks in the columnar (Col) phase (d).
 The color coding for the different phases is given in the inset, and white regions indicate the phase gaps between the coexisting phases.
 \textbf{(B)} Calculated phase diagram in terms of the attraction strength $u$ between the end groups of the semi-flexible rod-like particles as a function of their volume fraction $\phi$. 
 The different phases from isotropic liquid to the crystalline (Cr) state have been identified using global order parameters, as described in \cite{Braaf2017}.}%
 \end{figure*}

\begin{figure}
\includegraphics[scale=.25]{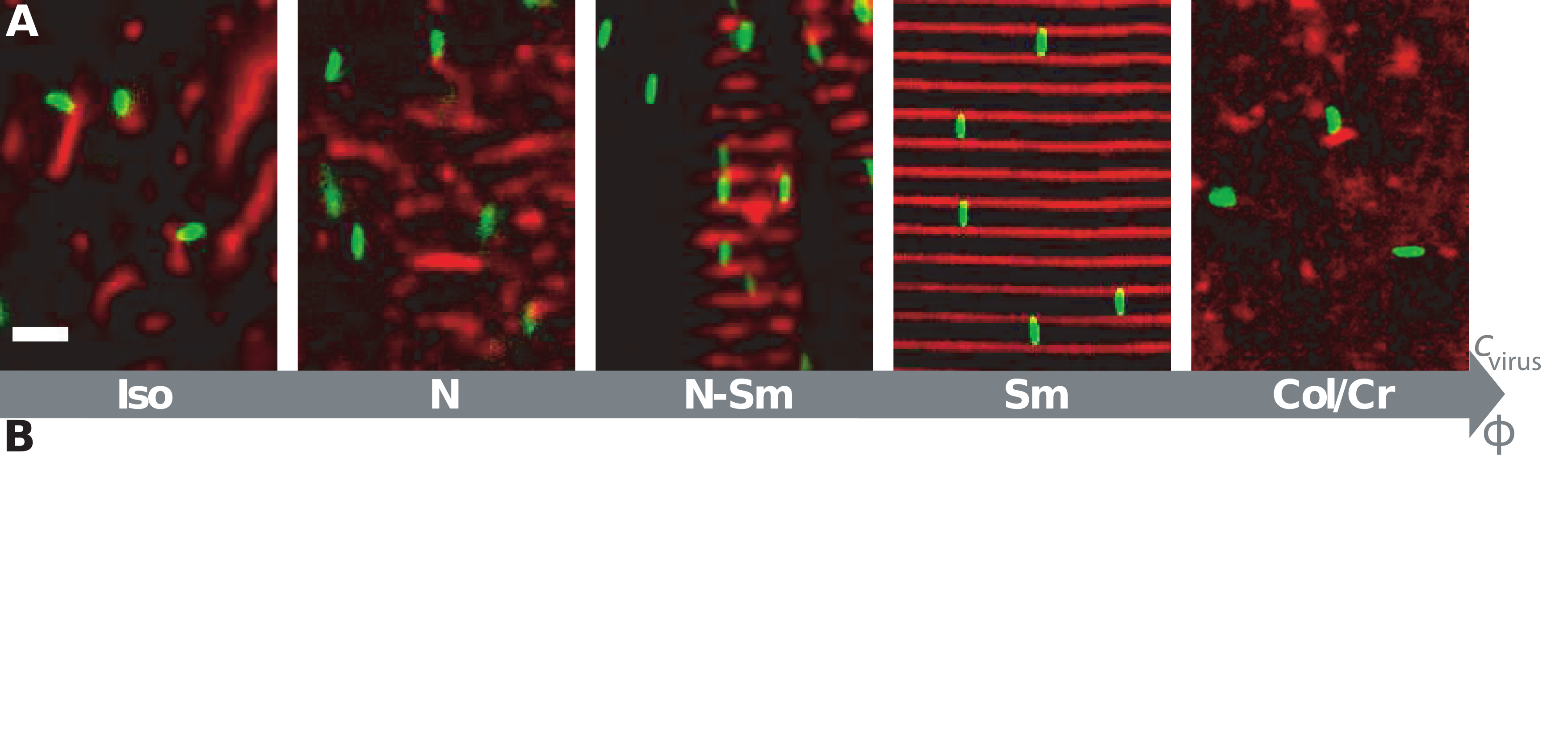}%
\caption{\label{texture}\textbf{(A)} Optical textures of tip-labeled virus suspensions in the various liquid crystalline phases obtained by fluorescence microscopy. The red signal corresponds to the position of the patchy virus tips. The green signal represent body-functionalized viruses, which are added in a tracer amount to the suspension. Scale bar: $2~\mu m$. \textbf{(B)} Corresponding simulation snapshots highlighting the local ordering within the mesophases. White beads are purely repulsive and red ones are attractive. In the isotropic phase (Iso), the viruses have random orientations. In the nematic phase (N), viral particles have an average alignment along the director but no long-range positional order. There are some clusters in both phases, which are interpreted as precursors of smectic layers. The first order nematic-smectic A phase transition (N-SmA) is confirmed by the presence of sharp interface between the coexisting phases. In the smectic phases (SmA and SmB), all patches are localized in the interlayer spacing. Conversely, in the columnar phase (Col), the virus patches are uniformly distributed. 
In this mesophase, body-functionalized viruses are not aligned due to small domain size. }%
 \end{figure}

We construct the experimental phase diagram for patchy rods as a function of the concentration and the degree of functionalization (Fig. \ref{phaseDiagram}A), and compare this with the phase behavior of the pristine viral particles. By comparing the stability limits of the various mesophases, which includes nematic, smectic A, smectic B and columnar phases, we conclude that increasing the number of dye molecules grafted at the tips of the virus particles strongly affects the nematic-smectic A (N-SmA) transition, yet has almost no effect over the other phase transitions. Our main finding is the increased stabilization of the smectic phase, at the expense of the nematic phase, with increasing number of grafted dyes, and concomitant widening phase gap implying that the transition becomes more strongly first order. 

Figure \ref{phaseDiagram}B presents our simulation phase diagram as a function of
the strength of the tip attraction, $u$. 
The resulting phase behavior shows qualitative agreement with the experimental data: increasing the stickiness of the tips affects mainly the nematic-smectic A phase transition. The stability of the smectic A phase increases with increasing strength of the tip attraction, as does the phase gap. 
For large enough attraction $u \gtrsim 1.8 k_BT$, we find in our simulations a direct isotropic liquid-to-smectic A phase transition, exploring a range of attraction that we cannot access experimentally due to the limited number of exposed cysteine groups at the virus tip (see the Supplemental Material \cite{SM}).
The isotropic liquid-to-nematic phase (I-N) transition remains unchanged in both phase diagrams, except for the highest tip attraction where the simulations point at a relatively weak widening of the coexistence range. 
This suggests that our patchy interaction is rather weak and localized, as rods with stronger attractive interaction, driven by either depletion interaction \cite{Tuinier,PRE2004} or by a residual van der Waals interactions between the bodies of the rods \cite{Lekkerkerker93}, 
exhibit a significant widening of the I-N coexistence range.

The results from our experiments and the simulations diverge at very high packing fractions. We do not find a stable columnar phase in our simulations of rod-like particles. This could be due to the difficulty of stabilizing the columnar organization in numerical simulations for entropy driven, single-component systems \cite{Dussi2018}. It is also possible that the columnar phase does not form in suspensions of particles with aspect ratios below 30, as suggested in \cite{Grelet2016}. Another obvious difference between experiments and simulations is the strongly first order transition between the smectic A and smectic B phases in the latter. Experimentally, it is second order or weakly first order \cite{Grelet2014}.
An extension of the smectic-B range by increasing the tip attraction that we find experimentally, is lacking in simulations for which there is also an intrinsic difficulty to clearly distinguish between the smectic B from the crystalline phase. 
The absence of one-to-one correspondence between the mass concentration in the experiments and the volume fraction in the simulations is not really surprising given the crude nature of the interaction potential, the modest aspect ratio of the particles in the simulations, and the overestimation of the size of the attractive bead in the simulations compared to the size of the attractive sites on the virus tip-proteins.

The overall qualitative agreement between experiments and simulations is however manifest. This is true for the dependence on tip attraction of the transitions between isotropic, nematic and smectic A phases (Fig. \ref{phaseDiagram}), but turns out to be true as well as for the local ordering displayed in these phases (Fig. \ref{texture}). For the purpose of direct comparison, we added a tracer amount of body labeled viruses with green fluorescent dyes to our suspensions. The striking feature of the optical texture as seen by fluorescence microscopy is the presence of red colored clusters in the isotropic phase.
By varying the depth of focus, we evidence the clusters to have a two-dimensional structure, forming bilayer ``lamellae" in which the viruses assemble at their red tips and lie nearly perpendicular to them. We cannot exclude the possibility that some of these clusters are caused by chemical rather than physical cross-linking, during the tip functionalization process.

Similar lamellar structures can be observed in the nematic phase, except that in this case they are oriented perpendicular to the director (defined as the average rod orientation) whereas in the isotropic phase they are randomly oriented (Fig. \ref{texture}A). Furthermore, the disappearance of the chiral nematic or cholesteric phase in favor of the uniaxial nematic phase upon grafting even a single dye molecule to the virus tip (Fig. \ref{phaseDiagram}A), we ascribe to the presence of these lamellae. We argue that they must interfere with the chirality amplification on the mesoscopic scale. In our simulations we observe bilayer clusters similar to those seen experimentally with particles assembled by their attractive tips in both sides, as shown by the snapshots in the isotropic and nematic phases displayed in Fig. \ref{texture}B.

At increased particle concentration, the lamellar aggregates grow and condense into smectic domains in a nematic background, corresponding to the N-SmA coexistence region
(See Fig. \ref{texture}, central images).
As expected, the particles are aligned along the director in the two phases, in both experiments and simulations. An example of the single smectic domain is given in Fig. \ref{texture}A, where the alignment of the rod-like particles is perpendicular to the layer allowing us to rule out any smectic C or other types of tilted smectic.

In contrast to the smectic A and B phases, which we are able to distinguish by means of SAXS measurements (see the Supplemental Material \cite{SM})\cite{Grelet2008,Grelet2014}, and which do exhibit large single domains, the columnar phase is characterized by finite domain sizes of only a few micrometers width, as shown both in Figs. \ref{phaseDiagram}A and  \ref{texture}A. The absence of bright red localized signals supports the lack of layered structure and is therefore consistent with the liquid-like order along the columns. The variation of red fluorescence intensity arguably does not reflect strong clustering, but may be interpreted as the result of the integration over the sample thickness of the fluorescence signal coming from domains with different orientations.

\begin{figure}
\includegraphics[scale=.25]{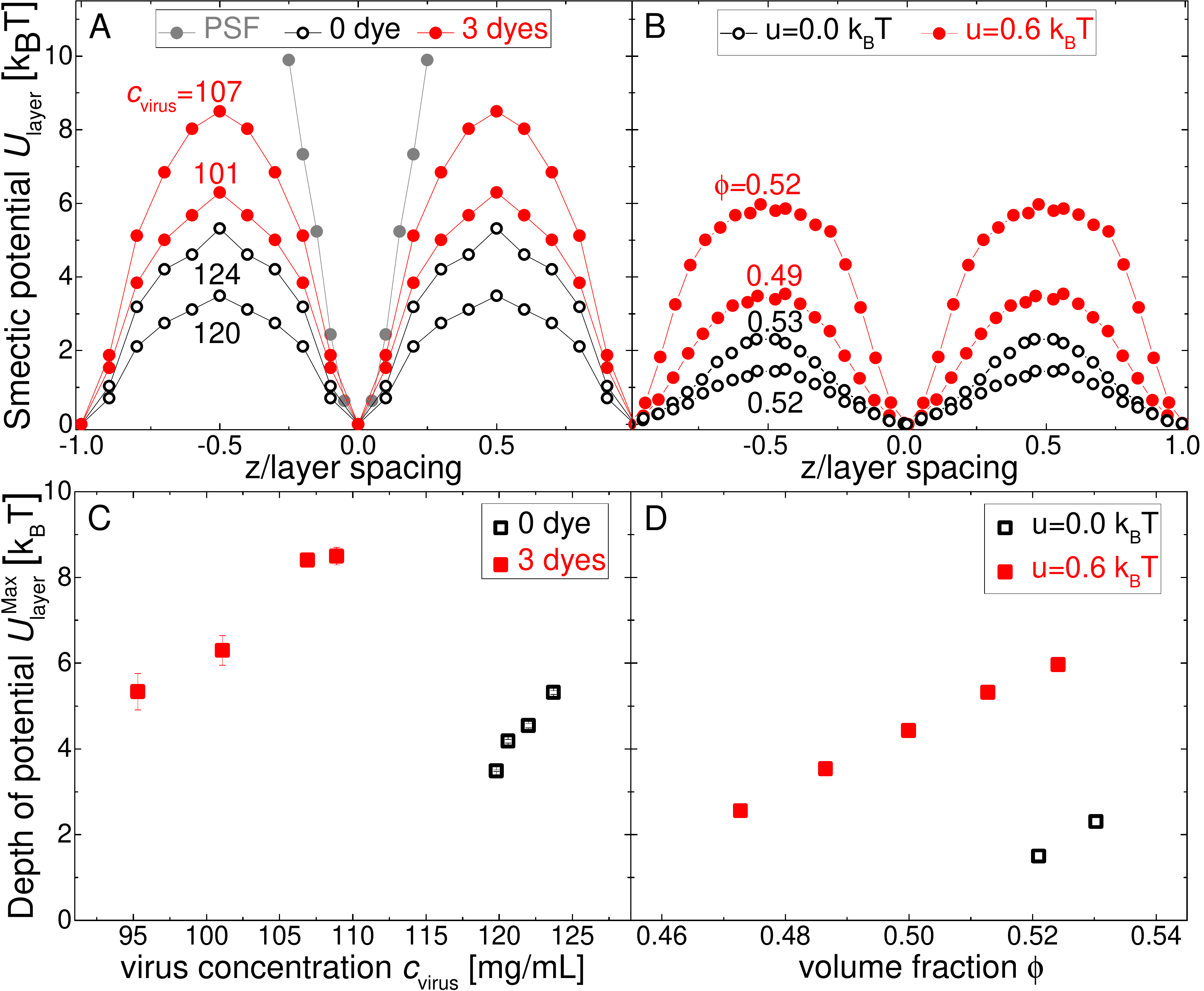}
 \caption{\label{smPotential} Smectic ordering potentials or molecular fields calculated from experimental \textbf{(A)} and simulated \textbf{(B)} distribution of particles, as a function of the particle position, normalized by the smectic layer spacing, for different particle concentrations and volume fractions. In gray we indicate the experimentally measured potential of immobile particles, which is the equivalent for potentials of the point spread function (PSF) of the optical setup. Smectic potential barriers as a function of virus concentration for experiments \textbf{(C)} and as a function of volume fraction for simulations \textbf{(D)}. In all graphs, open black and full red symbols correspond to ``raw'' repulsive and tip-functionalized rod-like particles, respectively.
}
\end{figure}

As the main effect of the tip patchiness is to widen the smectic stability range, we have characterized this phase by determining the associated molecular field $U_\mathrm{layer}$ \cite{Alvarez2017}. This unidimensional ordering potential can be obtained by measuring the distribution of longitudinal
rod fluctuations with respect to the middle of the layers, from which is deduced the probability $P(z)$ of finding a particle at position $z$ along the director. $P(z)$ is related to the ordering potential via the Boltzmann factor $P(z)\propto \exp (-U_\mathrm{layer}(z)/k_BT)$.
The free energy landscape of both experimental and simulated particles is presented in Fig. \ref{smPotential} and shows the same trends: (i) the magnitude of the ordering potential increases with increasing tip patchiness for a given particle packing fraction (Fig. \ref{smPotential}A-B), and (ii) $U_\mathrm{layer}$ increases with the particle concentration, for both repulsive and attractive tips (Fig. \ref{smPotential}C-D). Note in addition that the smectic potential also becomes narrower with increasing density and functionalizing the tips of the viruses. This implies that the amplitude of the fluctuations of the particles around their equilibrium positions in the layers become weaker, and hence that the particle positions become more localized. As the aspect ratio of the particles is smaller in our numerical simulations, we expect lower smectic potentials compared to the experimental ones, as shown in Fig. \ref{smPotential}. The reason is that the stability of the smectic A phase of repulsive rod-like particles reduces with decreasing length \cite{Bolhuis1997}. Notice that irrespective of the strength of tip attraction, we find the same slope of the ordering potential as a function of the particle concentration, both in the experiments and the simulations. This is to be expected because the molecular field a test particle experiences in a lyotropic smectic must be proportional to the average density \cite{Schoot1996}.
Even though we have not been able to find a sensible mapping between our experimental and simulation results because of the large disparity between the respective aspect ratios of the particles, our simulations do account for most of the features we observe in our experimental system. 
This is true both for the phase behavior and ordering potentials, suggesting that our prediction that a tip attraction strength as small as $u \approx 1-2~k_BT$ is sufficient to fully suppress the nematic phase and promote the smectic organization in dispersions of otherwise mutually repelling rod-like particles is plausible. This small value is actually not surprising, considering that free energy differences between particles in co-existing liquid crystalline phases of rod-like particles are typically of the order of a thermal energy and often much smaller than that.

In summary, we report on the achievement of tip-functionalized rod-like virus particles exhibiting sticky patches with tunable interaction. We find that the range of stability of the smectic phase of these particles can be enlarged continuously by increasing the strength of the patch attraction. Extending the stability of the smectic phase to lower concentrations happens at the expense of the nematic phase, in which bilayer lamellar aggregates form. Other phase transitions are, by and large, not affected by the tip functionalization. Our experiments and computer simulations suggest that the reason why only the smectic ordering responds to the tip functionalization, is that it brings together the interacting ends that are otherwise not highly correlated in the other phases.
Our findings open up perspectives in the rational design and site-specific post-modifications of 
particulate building blocks for soft self-assembled materials, showing how the introduction of a single and tiny enthalpic patch  is able to steer the structuring of complex fluids.

\begin{acknowledgments}
	This project has received funding from the European
	Union Horizon 2020 research and innovation programme
	under the Marie Sk\l{}odowska-Curie Grant Agreement No.
	641839. \\
\end{acknowledgments}

\bibliography{Rods}	

\clearpage

\end{document}